\documentstyle[amssymb,aps,multicol,epsf]{revtex}
\begin{document}
\title{Effective Second Order Susceptibility in Photonic Crystals Composed from
Centro-Symmetric Materials }
\author{A. Feigel, Z. Kotler and B. Sfez}
\address{Soreq NRC, Yavne 81800, Israel}
\maketitle

\begin{abstract}
A technique for obtaining efficient bulk second order susceptibility in non
centro-symmetric Photonic Crystals (PC) made from centro-symmetric materials
is discussed. The effect is based on the electric quadrupole effect, strong
electromagnetic mode deformation and non homogeneous contribution to volume
polarization from different parts of the PC. The required symmetry breaking
is introduced on the macroscale of the PC unit cell. The obtained structural 
$\chi _{\rm str}^{\left( 2\right) }$ is comparable with the second order
susceptibility of ordinary non-linear materials. Phase matching can be
achieved by introducing symmetry modulation (Quasi Phase Matching) during
fabrication of the PC.
\end{abstract}

\begin{multicols}{2}

Optical materials with high second order susceptibility $\chi ^{\left(
2\right) }$ are frequently required for both fundamental and applied
research. By utilizing three wave interactions in such media, existing
coherent radiation sources can be extended to almost the entire optical
spectrum. Consequently the importance of these materials can be compared
with the importance of laser itself.

The number of efficient second order nonlinear materials is limited. To
possess second order susceptibility the material has to be
non-centro-symmetric. This immediately eliminates all amorphous materials
and crystals from 11 of 32 symmetry classes. The choice between remaining
materials is also limited due to different constraints: value of $\chi
^{\left( 2\right) }$, absorption in required spectrum, damage threshold etc.

Different methods were proposed for development of enhanced non-linearity.
The first is the search for new materials with large non-linear response on
molecular or molecular arrangement\cite{DyeS} scales. The second is
construction of composite materials\cite{boyd}. In this case the effective
enhancement is achieved due to redistribution of the energy inside composite
structure. Unfortunately it can improve only $\chi ^{\left( 3\right) }$, and
not $\chi ^{\left( 2\right) }$. Recently the use of Photonic Band Gap (PBG)
crystals made from non-linear materials was proposed\cite{Bowden},\cite
{johnnon}. All these methods require symmetry breaking on microscopic atomic
or molecular dimensions.

In this Letter we describe a technique for obtaining efficient second order
susceptibility in non centro-symmetric Photonic Crystals (PC) made from
centro-symmetric materials. The effect is based on the electric quadrupole
transition, strong electromagnetic mode deformation and different
contributions to the volume polarization from different parts of the PC. The
required symmetry breaking is introduced on the macroscale of the PC unit
cell.

A local second order polarization $P^{\left( 2\right) }$ can be obtained in
centro-symmetric materials due to the quadrupole effect\cite{bloem}. In
quadrupole transitions the required symmetry breaking is obtained by
asymmetry of the electromagnetic field spatial mode, rather than the
asymmetry of the electron wave function as in the dipole transition case.
The second order polarization corresponding to a quadrupole transition is: 
\begin{equation}
\overrightarrow{P}_{Q}^{(2)}=Q\vdots \overrightarrow{E}\nabla 
\overrightarrow{E}  \label{eq1}
\end{equation}
where $Q$ is a fourth-order tensor. Generally integration over the volume of
such point polarization vanishes, due to periodicity of $\overrightarrow{E}$
and gradient dependence of $P_{Q}^{\left( 2\right) }$. Usually, only some
weak signals from interfaces can be detected\cite{sion},\cite{Dadap},
however the situation in PCs can be quite different.

PCs are artificial two or three dimensional periodic dielectric structures 
\cite{Eli},\cite{jhon}. They possess unique optical properties for
electromagnetic radiation with wavelengths comparable with their period,
including the existence of full PBGs, anomalously strong dispersion\cite
{superprism} and high photon localization near defects \cite{Joan}. High
dielectric constant modulation is required to obtain strong effects, so
several technologies for construction of dielectric/air 2D \cite{2dscherer}, 
\cite{rue} and 3D\cite{Lin}, \cite{Noda}, \cite{Soreq} PCs have been
developed.

Integration of eq. (\ref{eq1}) over the volume in dielectric/air PC can be
different from zero. The reason is unequal contributions to the polarization
from different parts of the media. The polarization of the air regions can
be totally neglected due to low electron density. Constructing PC in such a
way that in dielectric part the $P_{Q}^{\left( 2\right) }$ has one sign and
in the air the opposite, effective ''structural'' volume polarization can be
obtained.

\begin{figure}[h]
  \begin{center}
    \epsfxsize=2.8in
    \hskip0.15in\epsfbox{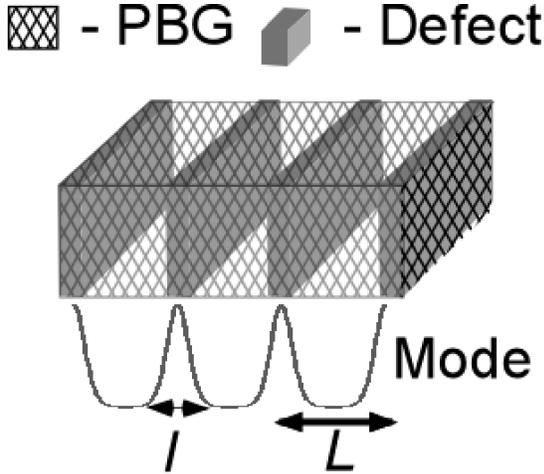}
  \end{center}
\caption{1D defect lattice, with lattice constant $L$, in Photonic
Band Gap (PBG) environment. A single defect in PBG environments can possess
localized modes for frequencies inside the gap. This means that for the same
frequencies in periodic lattices of defects the mode can be not localized,
but strongly modulated. $l$ is the characteristic modulation length.}
\label{FIG1}
\end{figure}

Effective second order susceptibility $\chi _{Q}^{\left( 2\right) }$ induced
by a quadrupole effect can be quite large \cite{bloem}:
\begin{equation}
\frac{\chi _{Q}^{\left( 2\right) }}{\chi ^{\left( 2\right) }}\approx \frac{d}
{\lambda }\eta   \label{ratio}
\end{equation}
where $\chi_{Q}^{\left( 2\right) }\approx \left| Q\right| /l$, $\lambda $
is the radiation wavelength, $\chi ^{\left( 2\right) }$ is some
characteristic second order susceptibility, $d$ is characteristic
interatomic dimension, $l$ is the modulation scale of electromagnetic mode
(see Fig.~\ref{FIG1}) and $\eta $ is some numerical coefficient. $l$ can
be as small
as the radiation wavelength $\lambda $ and $d\approx 0.3nm$. The origin of
$\eta \geq 10$ is that the dipole transition matrix element between states of
different parity always has smaller numerical coefficient than quadrupole
transition based on the matrix element of the same parity\cite{bloem}. The
experiments on the surface non-linear effect in Si, high index material
which is used for PCs fabrication, show that in this material $\chi
_{Q}^{\left( 2\right) }$ is only about three orders of magnitude weaker than 
$\chi ^{\left( 2\right) }$ of GaAs for $\lambda \approx 1\mu m$\cite{lee}.
It means that properly designed PCs made from Si, with non-vanishing
contribution of $\overrightarrow{P}_{Q}^{(2)}$ in all the volume, can
artificially provide second order susceptibility at least of the same order
of magnitude as low second order coefficient uniform materials (e.g.
quartz). The result may be even larger due to stronger modulation of the
electromagnetic field in PC than at an ordinary interface and due to the
influence of photonic band structure on electron transitions
\cite{johnnon}.

A specific realization of the desired material requires PCs with highly
modulated electromagnetic mode. It can be achieved directly in some PCs,
e.g. high mode modulation was predicted for silicon inverse opal structure 
\cite{jhonmode}, or by introducing periodic defects into the PBG substrate.
The first approach may be realized using lower index materials, because the
requirement for PBG does not exist. Unfortunately there were no intensive
numerical search for photonic structures with high mode modulations and at
these moment PCs of this type are not well explored. In contrast PBG
crystals require high refractive index materials, but there are many known
structures and almost any defect (extra or missing material, deviation from
the periodicity) possesses localized modes for frequencies inside the gap,
leading to modulated mode structure (see Fig.~\ref{FIG1}).

Also in the latter approach the required asymmetry can be achieved by proper
defect's design. Independent of the PBG substrate the defects should be
asymmetric and possess dielectric/air structure. In the case when the
electromagnetic mode is omnidirectional in dielectric part of the defect,
the maximum volume contribution of polarization eq. (\ref{eq1}) can be
obtained. In our opinion, the hollow cavities, partially filled with the
substrate's material, are the best candidates for proposed method.

\begin{figure}[h]
  \begin{center}
    \epsfxsize=2.8in
    \hskip0.15in\epsfbox{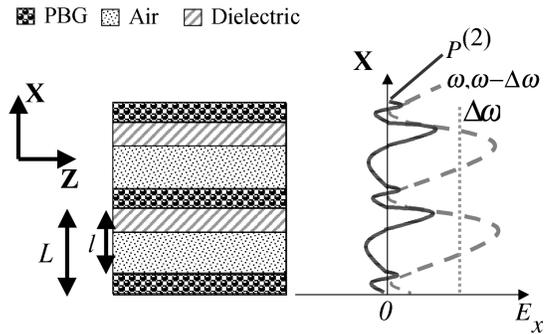}
  \end{center}
\caption{Periodic lattice of waveguides (the size of each waveguide is 
$l\approx 2\pi c/\omega $) in 2D PBG crystal, and the TE electromagnetic
mode structure inside it. The mode is modulated due to the strong decay of
electromagnetic field in PBG crystals. The photonic crystal (PC) is non
centro-symmetric due to partially dielectric filling of waveguides. Guiding
waves in such waveguides is an intrinsic property of PBG crystals. The
figure is a top view of the proposed 2D photonic crysal and can be
considered as an image of the lithography mask, required for its
fabrication.}
  \label{FIG2}
\end{figure}

Let us now consider some specific example (see Fig.~\ref{FIG2}): Optical
Parametric Oscillations (OPO) in a lattice of 1D defects (hollow
waveguides) in 2D PBG environment. Such waveguides array can be fabricated
using semiconductor lithography technology. To obtain a 2D PBG crystal, a
hexagonal pattern of deep air columns can be etched in the high refractive
index substrate, e.g. Si. During the same lithographic step the waveguides
can be introduced by etching parallel stripes. Using quasicrystalline
arrangement of air columns instead of hexagonal pattern, 2D PBG can be
achieved in lower index materials, e.g. glass\cite{glass}.

Inversion symmetry breaking is achieved by partial waveguide filling. It can
be done by leaving an unetched stripe near each etched one(see
Fig.~\ref{FIG2}). The
possibility to guide electromagnetic modes in such waveguides is intrinsic
property of PBG crystals. Guiding in PBG crystals has no crucial dependence
on the internal waveguide's structure, as in index guided modes, because the
confinement of the electromagnetic mode exists only due to the external PBG
environment.

Three electromagnetic waves (each TE mode): pump, signal and idler,
propagate along the waveguides directions. Let us assume that pump and
signal have almost degenerate frequencies $\omega $ and $\omega -\Delta
\omega $ inside the forbidden gap. The idler's frequency $\Delta \omega $ is
assumed to be outside the gap (see Fig.~\ref{FIG2}).

Propagation of some specific radiation in a non-linear waveguide, can be
described under slow varying amplitude approximation as\cite{Shen}: 
\begin{eqnarray}
\frac{\partial A}{\partial z} &=&\frac{2\pi i 
\omega^2}{Kc^{2}}\frac{1}{\sqrt{D}}\int d^{2}\rho \,
\overrightarrow{F}(\rho)
\overrightarrow{P}^{NL}(\rho) \, e^{-iKz+i\omega t},
\label{eq8} \\
D &=&\int d^{2}\rho \, \overrightarrow{F}(\rho)
\overrightarrow{F}(\rho) \nonumber
\end{eqnarray}
where $A$ is an amplitude ($E=AF(\rho )/\sqrt{D}\exp (iKz-i\omega t)$),
$\omega$ and $K$ are the frequency and wavevector of the mode, $c$ is the
speed of light, $\overrightarrow{F}\left( \rho \right) $ is the normalized
field distribution in the transverse plane, $\rho $ is transverse coordinate
and $\overrightarrow{P}^{NL}\left( \rho \right) $ is the induced non-linear
polarization. For our specific model (see Fig.~\ref{FIG2}) the equation
for the
signal eq. (\ref{eq8}) can be written as: 
\begin{eqnarray}
\frac{\partial A^{(s)}}{\partial z} &=& \frac{2\pi i 
\omega_{s}^{2}}{K_{s}c^{2}}\frac{1}{\sqrt{D_{s}}}\int dx \, F_{x}^{(s)}
\left( x\right) |Q| \cdot \nonumber \\
 && {} \cdot E_{x}^{\left( i\right) \ast }
\frac{\partial E_{x}^{\left( p\right) }\left( x\right) }{\partial x}
\, e^{-i\Delta kz},   \label{eq9} \\
\Delta k &=&K_{p}-K_{s}-K_{i}  \nonumber
\end{eqnarray}
using eq. (1) for $\overrightarrow{P}^{NL}$ and that the field of TE modes
has only $x$ direction.

Under the assumption that the idler wavelength is larger than the defect's
size, $E_{x}^{\left( i\right) }$ can be assumed to be constant in the
transverse plane. Under the same assumption the modes of pump and signal can
be taken to be identical with some maximum $E_{\rm max}^{\left( p\right)
}$on
the dielectric/air interface inside the waveguide and zeros on its boundaries
\cite{Joan},\cite{scherer} (see Fig.~\ref{FIG2}). In this case the
integral in eq.
(\ref{eq9}) does not depend on the exact electromagnetic mode distribution,
but on the field value at the boundary points of integration. The integral
is tacken only over the filled part of the waveguide and can be evaluated
analytically: 
\begin{equation}
\frac{\partial }{\partial z}A^{\left( s\right) }=\frac{2\pi i \omega
_{s}^{2}}
{K_{s}c^{2}}\frac{\left| Q\right| }{\sqrt{l}}E^{\left( i\right) \ast
}E_{\rm max}^{\left( p\right) } \, e^{-i\Delta kz}.
\label{eq10}
\end{equation}
For an ordinary $\chi^{\left( 2\right) }$ process eq. (\ref{eq9}) becomes: 
\begin{equation}
\frac{\partial }{\partial z}A_{\rm mean}^{\left( s\right) }=\frac{2\pi i
\omega_{s}^{2}}{K_{s}c^{2}}\sqrt{L}\chi^{\left( 2\right) }E^{\left(
i\right) \ast
}E_{\rm mean}^{\left( p\right) } \, e^{-i\Delta kz}.
\label{oponorm}
\end{equation}
To compare eqs. (\ref{eq10}) and (\ref{oponorm}) one must take into
account
that the energies of pump modes should be the same in both cases: 
\begin{equation}
E_{\rm max}^{2}l\approx E_{\rm mean}^{2}L.  \label{moderel}
\end{equation}
The ratio of effective non-linearities in these processes is: 
\begin{equation}
\frac{\chi _{\rm str}^{\left( 2\right) }}{\chi ^{\left( 2\right)
}}=\frac{\left|
Q\right| }{\chi ^{\left( 2\right) }}\frac{1}{l}\sqrt{\frac{l}{L}}.
\label{eq12}
\end{equation}
It can be rewritten using eq. (\ref{ratio} ) and that $l\approx \lambda $
as: 
\begin{equation}
\frac{\chi_{\rm str}^{\left( 2\right) }}{\chi ^{\left( 2\right)
}}=\frac{d}
{\lambda }\eta \beta_{\rm overlap}  \label{opores}
\end{equation}
where $\beta_{\rm overlap}=\sqrt{l/L}$ indicates imperfect modes overlap
in eq.
(\ref{eq8}). It can not affect the process seriously, since $L$ and $l$ can
be of the same order ( $L\approx 2l$ is realistic) due to strong light
confinement in PBG crystals, hence previously obtained estimations are valid
for this specific PC.

For efficient conversion phase matching has to occur, $\Delta k=0$ in eq.
(\ref{eq9}). Otherwise the signal from different points along the propagation
will interfere destructively. Generally in most materials this condition can
be influenced only by change of the propagation direction.

\begin{figure}[h]
  \begin{center}
    \epsfxsize=2.8in
    \hskip0.15in\epsfbox{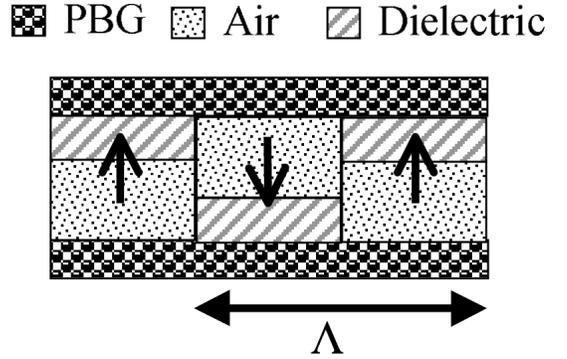}
  \end{center}
\caption{Realization of Quasi Phase Matching in PCs for second order
non-linear processes, $2\pi /\Lambda =\Delta k$. The phase matching is
achieved by period modulation of the asymmetry's direction. It can be done
by mask modification during lithography step. In 3D the task can be
completed by lithography multilayer construction. The desired asymmetrically
filled cavity can be constructed by changing the defect's image from
negative to positive in different layers.}
  \label{FIG3}
\end{figure}

Phase matching can be achieved artificially by periodic modulation of the
sign of the non-linear tensor coefficient. It can be achieved in
ferroelectric crystals only by periodic poling. The method is called Quasi
Phase Matching \cite{arm}. In non centro-symmetric PCs modulation of the
non-linear coefficient can be introduced during fabrication (see
Fig.~\ref{FIG3}).

These results can be extended to the general case of periodic lattice
defects inside a PBG Crystal. As it was stated earlier single defects in PBG
environments can possess localized modes for frequencies inside the gap.
This means that for the same frequencies in periodic lattices of defects the
mode can be not localized, but strongly modulated. Assuming that the
electric field near the defects is $E_{\rm max}$ and between them is
$E_{\min
}\approx 0$, due to eq. (\ref{eq1}) the scaling for structural polarization
is: 
\begin{equation}
P_{\rm str}^{\left( 2\right) }\propto \left| Q\right| E_{\rm max}^{2}l^{N-1}
\label{eq3}
\end{equation}
where $l$ is the effective localization length of the mode and $N$ is the
dimension of the defect's lattice. This result is valid for the optimal case
of field distribution, e.g. as in the previously considered example. The
dimension of the defect lattice can be smaller than the space dimension. The
periodicity of the entire crystal implies translational symmetry along all
unconfined coordinates. Hence these coordinates can be omitted (see
Fig.~\ref{FIG1}).

The ''structural'' polarization has to be compared with the ordinary one: 
\begin{equation}
P^{\left( 2\right) }\propto \chi ^{\left( 2\right) }E_{\rm mean}^{2}L^{N}
\label{eq4}
\end{equation}
where $L$ is the defect lattice constant. The ratio between $E_{\rm max}$
and
$E_{\rm mean}$ can be determined taking into account that the total
energies of
the compared modes has to be equal, similarly to eq. (\ref{moderel}): 
\begin{equation}
\alpha E_{\rm max}^{2}l^{N}=E_{\rm mean}^{2}L^{N}  \label{eq5}
\end{equation}
where $\alpha $ is some numerical coefficient of order $1$. Taking into
account eq. (\ref{ratio}) one can get: 
\begin{equation}
\frac{P_{\rm str}^{\left( 2\right) }}{P^{\left( 2\right) }}=\frac{\chi
_{\rm str}^{\left( 2\right) }}{\chi ^{\left( 2\right) }}=\frac{d}{l}\eta
\beta_{\rm overlap}.  \label{eq7}
\end{equation}

The additional numerical coefficient $\beta_{\rm overlap}$ corresponds to
imperfect overlap of different modes \cite{yariv}. In case of three wave
interactions its scaling can be determined from eqs.
(\ref{eq3}, \ref{eq4}, \ref{eq5}) to be: 
\begin{equation}
\beta_{\rm overlap}\approx 1
\end{equation}
in the case of all waves being confined. Practically this can be difficult
due to the finite forbidden gap in PCs. In case only two waves are confined
$P_{\rm str}^{\left( 2\right) }\propto \left| Q\right| E_{\rm mean}E_{\rm
max}
l^{N-1}$
and the scaling is: 
\begin{equation}
\beta _{\rm overlap}\varpropto (l/L)^{N/2}
\label{betascal}
\end{equation}
where $N$ is the dimension of the defect's lattice. For $N=1$ results (\ref
{eq7}) and (\ref{opores}) are equivalent. The obtained results are not
limited to mode modulation by defects lattice, but the same formulae can be
applied in the case of ''pure'' PC strong mode modulation \cite{jhonmode}.

PC fibers\cite{fiber} and so called ''Super Mirror''\cite{mirror} waveguides
are other good candidates for implementation of second order non-linear
processes based on an electric quadrupole transition. They are suited for
guiding light mode in the air, consequently the partial filling of these
waveguides can provide effective second order non-linearity. Recently very
interesting non-linear processes were observed in air/silica microstructure
optical fibers\cite{Ranka}. Three dimensional defects are also highly
interesting due to the analogy between 3D defects and molecules. All these
cases require more extensive numerical investigation.

The measurement of the effects value can be completed by light propagation
inside a properly designed PC, e.g. in the proposed waveguide's array
structure, or by light scattering from the surface of PC. In this case bulk
and surface non-linearities can be distinguished by contribution of
different light polarizations to the signal. In the case of surface
non-linearity of centro-symmetric materials the tangential component of
electromagnetic field does not contribute to generated harmonics of incident
light.

Integrated optics is one of the possible applications for structural $\chi
_{\rm str}^{\left( 2\right) }$ materials. Use of standard non-linear
materials
in integrated optics is difficult. Generally integration is either
impossible due to incompatibility of different processes or expensive. The
most convenient are polymers, but several specific disadvantages (e.g.
lifetime) prevent their broad use. The method described above opens new
possibilities for incorporating non-linear elements in optical chips by e.g.
obtaining effective non-linearity from the chip substrate itself.

It was shown that a non centro-symmetric PC, made from substrates without
bulk second order nonlinearity (amorphous materials or centrosymmetric
crystals) can posses effective volume $\chi ^{\left( 2\right) }$ comparable
with the second order susceptibility of ordinary nonlinear materials. The
effect is based on the electric quadrupole transition and local enhancement
of $\nabla \overrightarrow{E}$. The symmetry breaking is introduced on the
macroscale of the unit cell of the PC, instead of atomic scale. The
estimations were made for $\lambda =1\mu m$, and for smaller wavelengths the
non-linearity should be stronger. Quasi Phase Matching can be introduced in
the time of fabrication, so effective non-linear processes are feasible.

Discussions with Dr. G. Berkovich and Prof. Y. Silberberg are gratefully
acknowledged.

\end{multicols}

\end{document}